\documentclass[%
reprint,
superscriptaddress,
nofootinbib,
amsmath,
amssymb,
aps,
prc,
floatfix,
showkeys,
]{revtex4-2}

\usepackage{graphicx}
\usepackage{dcolumn}
\usepackage{bm}
\usepackage[normalem]{ulem}
\usepackage{fancyhdr}
\usepackage{graphicx,amsfonts,amssymb,amsbsy}
\usepackage{amsmath,amsthm,latexsym}
\usepackage[utf8]{inputenc}  
\usepackage{natbib}
\usepackage[colorlinks]{hyperref}
\usepackage{booktabs}
\usepackage{aas_macros}
\usepackage{xcolor}
\usepackage{xspace}
\usepackage{soul}

\newcommand{\soft}{{Soft}\xspace}
\newcommand{\inter}{{Intermediate}\xspace}
\newcommand{\stiff}{{Stiff}\xspace}

\begin{document}

\title{Could a slow stable hybrid star explain the central compact object in HESS~J1731-347?}

\author{Mauro Mariani}
\email{mmariani@fcaglp.unlp.edu.ar}
\affiliation{Grupo de Astrofísica de Remanentes Compactos, Facultad de Ciencias Astronómicas y Geofísicas, Universidad Nacional de La Plata, Paseo del Bosque S/N, La Plata (1900), Argentina}
\affiliation{CONICET, Godoy Cruz 2290, Buenos Aires (1425), Argentina}

\author{Ignacio F. Ranea-Sandoval}
\email{iranea@fcaglp.unlp.edu.ar}
\affiliation{Grupo de Astrofísica de Remanentes Compactos, Facultad de Ciencias Astronómicas y Geofísicas, Universidad Nacional de La Plata, Paseo del Bosque S/N, La Plata (1900), Argentina}
\affiliation{CONICET, Godoy Cruz 2290, Buenos Aires (1425), Argentina}

\author{Germán Lugones}
\email{german.lugones@ufabc.edu.br}
\affiliation{Universidade Federal do ABC, Centro de Ciências Naturais e Humanas, Avenida dos Estados 5001- Bangú, CEP 09210-580, Santo André, SP, Brazil.}

\author{Milva G. Orsaria}
\email{morsaria@fcaglp.unlp.edu.ar}
\affiliation{Grupo de Astrofísica de Remanentes Compactos, Facultad de Ciencias Astronómicas y Geofísicas, Universidad Nacional de La Plata, Paseo del Bosque S/N, La Plata (1900), Argentina}
\affiliation{CONICET, Godoy Cruz 2290, Buenos Aires (1425), Argentina} 

\begin{abstract}
We explore an alternative explanation for the low-mass ultra-compact star in the supernova remnant HESS~J1731-347 using a model-agnostic approach to construct hybrid equations of state. The hadronic part of the hybrid equation of state is constructed using a generalized piecewise polytropic scheme, while the quark phase is described by the generic constant speed of sound model. We assume an abrupt first-order hadron-quark phase transition with a slow conversion speed between phases. Our equations of state align with modern Chiral Effective Field Theory calculations near nuclear saturation density and are consistent with perturbative Quantum Chromodynamics calculations at high densities. Using this theoretical framework, we derive a wide range of hybrid equations of state capable of explaining the light compact object in HESS~J1731-347 in a model-independent manner, without fine-tuning. These equations of state are also consistent with modern astronomical constraints from high-mass pulsar timing, NICER observations, and multimessenger astronomy involving gravitational waves. Our results support the hypothesis that the compact object in HESS~J1731-347 could plausibly be a slow stable hybrid star.
\end{abstract}

\maketitle

\section{Introduction}
\label{sec:intro}

The lightest neutron star (NS) confirmed to date, with a mass of $M = 1.174 \pm 0.004\,M_{\odot}$, is located within a double pulsar system known as J0453+1559 \cite{Martinez:2015pja, Ozel:2016mra} (although the authors of Ref.~\cite{tauris:2019jan} propose that the lower-mass component of J0453+1559 might be a white dwarf instead of a NS). However, the recent discovery of a potentially even lighter NS poses a new challenge to astrophysics in the quest for understanding the equation of state (EoS) for dense matter. In a recent paper, \citet{Doroshenko:2022asl} estimated, after modeling the X-ray spectrum and using a Gaia-observation-based distance, the mass, $M$, and radius, $R$, of the central compact object (CCO) in the supernova remnant HESS~J1731-347. The best values obtained for these quantities are $M = 0.77^{+0.20}_{-0.17}\,M_\odot$ and $R=10.4^{+0.86}_{-0.78}$~km, respectively, at the $1\sigma$ posterior credible levels.  

It is important to recall, however, that several assumptions are needed to obtain such tight results. In particular, it is assumed that the atmosphere of the NS is predominantly composed of carbon and that it has a uniform temperature distribution; such hypothesis, leading to a strong model-dependence, is subject to criticism and discussion. Indeed, recent studies claim that most known CCOs do not exhibit uniform-temperature carbon atmospheres \cite{Alford:2023dcc}. Instead, they would show evidence of nonuniform temperature surfaces with small hot spots, suggesting that uniform-temperature carbon atmospheres might not be a common characteristic among observed CCOs, and instead, multitemperature surfaces are more typical. In fact, the analysis made by \citet{Alford:2023dcc} of the longest, highest-quality data set from XMM-Newton produces results inconsistent with a uniform-temperature carbon atmosphere model, as suggested by \citet{Doroshenko:2022asl}, whether or not attempts are made to model dust scattering. \citet{Doroshenko:2022asl} also considered other viable assumptions and the results obtained for the mass and radius of the CCO are different. Ref.~\cite{Alford:2023dcc} also highlights another key aspect: the assumption made by \citet{Doroshenko:2022asl} regarding the distance at which the CCO is located is crucial. Their use of a distance estimate of $2.5 \, \mathrm{kpc}$, not supported by other studies, is critical for concluding that the mass of the CCO in HESS J1731-347 is significantly lower than that of a canonical NS. It is also important to note that such estimation is in tension with numerical simulations of supernovae explosions, which suggest that the minimum possible mass for the remnant NS has to be larger (see, for example, Ref.~\cite{Suwa:2018otm} and references therein). Nevertheless, if these values are confirmed by future observations and more detailed analysis, such object would be one of the least massive and smallest compact objects ever observed. Hence, it represents a relevant and challenging observation to analyze and elucidate.

In addition to these low-mass compact objects, there are other astronomically relevant estimations of mass and radius from double pulsars \cite{Demorest:2010sdm,Antoniadis:2013amp,Arzoumanian:2018tny, Cromartie:2020rsd, Fonseca:2021rfa}, multimessenger astronomy with gravitational waves from event GW170817 \cite{Abbott:2017gwa,Abbott:2017oog,Abbott:2018gmo,Abbott:2018pot} and NICER observations of the isolated NS PSR~J0030+0451 \cite{Miller:2019pjm,Riley:2019anv} and the double pulsar PSR~J0740+6620 \cite{Miller:2021tro,Riley:2021anv}. Current dense matter EoS and compact object models should simultaneously address all these astrophysical constraints, along with those from nuclear and particle data. Regarding this latter family of restrictions, there are constraints for the dense matter EoS in the two limit regimes of low and high density. For densities ${n} \lesssim n_0$, with $n_0=0.16~$fm$^{-3}$ the nuclear saturation density, the EoS is determined with accuracy due to nuclear theory and experiments. For the range $n_0 \lesssim n \lesssim 2n_0$, there are calculations based on the chiral Effective Field Theory (cEFT). Recently, various improved cEFT calculations have been published  \cite{Hebeler:2013nza, Lynn:2016ctn, Hu:2017nmp, Holt:2017eos, Drischler:2020hwd}; in our work, we use the calculations provided by \citet{Drischler:2021lma}, who obtained a refined constraint for $\beta$-stable NS matter up to $2 n_0$ using a Bayesian framework for quantifying and propagating correlated cEFT truncation errors. For very high densities, in the $n \gtrsim 40\,n_0$ regime, perturbative QCD (pQCD) calculations help understand the behavior of deconfined matter and derive the corresponding EoS \cite{Gorda:2018gpy}. As we will discuss later, such extreme densities, although unusual or absent in traditional hadronic NSs, could be relevant in some exotic scenarios \cite{lugones:2023ama}.

According to the proposals in the literature, NSs constructed with soft EoS can explain the low-mass object observed in HESS~J1731-347 (see, for example, Refs.~\cite{Sagun:2023wit, Kubis:2023rmf} and references therein for modern studies on this scenario). In addition to these models, other exotic theoretical scenarios have been proposed to explain such a low-mass compact object. In Refs.~\cite{Laskos:2024hsi, Sagun:2023wit}, the authors analyzed the possibility that this compact object might be a hybrid star (HS). Furthermore, more radical theoretical possibilities have been explored: for example, it has been proposed that the HESS~J1731-347 object might be a strange quark star (SQS) (see, for example, Refs.~\cite{DiClemente:2022-arXiv-itc, Horvath:2023als, das:2023ass, Lugones:2023eve, Sagun:2023wit, Lugones:2024ryz} and references therein) or a compact object in which normal matter is admixed with dark matter \cite{Sagun:2023wit}.

In view of the wide variety of theoretical alternatives proposed in the literature regarding the nature of the compact object at the center of the HESS~J1731-347 supernova remnant, we aim to explore the possibility that the low-mass compact object could be a \textit{slow stable} hybrid star (SSHS), a concept previously introduced and discussed in several papers \cite{Pereira:2017rmp, mariani:2019mhs, rodriguez:2021hsw,curin:2021hsw, Mariani:2022omh, Goncalves:2022ios, Ranea:2022bou, Ranea:2023cmr, Ranea:2023auq, Rau:2023tfo, lugones:2023ama, Gosh:2024ero, Rather:2024roo}. SSHSs are compact stars with a sharp hadron-quark phase transition that could remain stable even though $\partial M /\partial \epsilon_c < 0$, when considering a \textit{slow} hadron-quark conversion speed such that the conversion timescale is larger than the typical oscillation timescale. Within this theoretical possibility, perturbed fluid elements at the phase-transition surface do not have enough time to convert quarks into hadrons (or vice-versa). For this reason, motion around the interface solely involves stretching and compression of volume elements but with no transformation between the different phases. This feature differentiates these configurations from the more traditional totally stable hybrid stars (TSHSs), which are located in branches that satisfy $\partial M /\partial \epsilon_c > 0$ and are stable under both \textit{slow} and \textit{rapid} conversion scenarios. The significance of SSHSs lies in their potential to explore extreme regions of the QCD phase diagram through astronomical observations of compact objects. Moreover, as stated in Ref.~\cite{lugones:2023ama}, ultra-stiff EoS should not be discarded, as astronomical constraints (such as those from GW170817 and its electromagnetic counterpart, along with the radius measurement of PSR J0740+6620) can be satisfied by SSHSs.

The paper is organized as follows. In Section~\ref{sec:models} we describe the parametric model used to construct the hybrid EoS and the role of sharp and slow hadron-quark phase transitions in the stability analysis of HSs. Section~\ref{sec:results} presents the most important results of our study, while their astronomical implications are discussed in Section~\ref{sec:conclus}.

\section{Hybrid EoS Construction Criteria}
\label{sec:models}

\begin{figure*}[tbh]
\centering
\includegraphics[height=.353\linewidth]{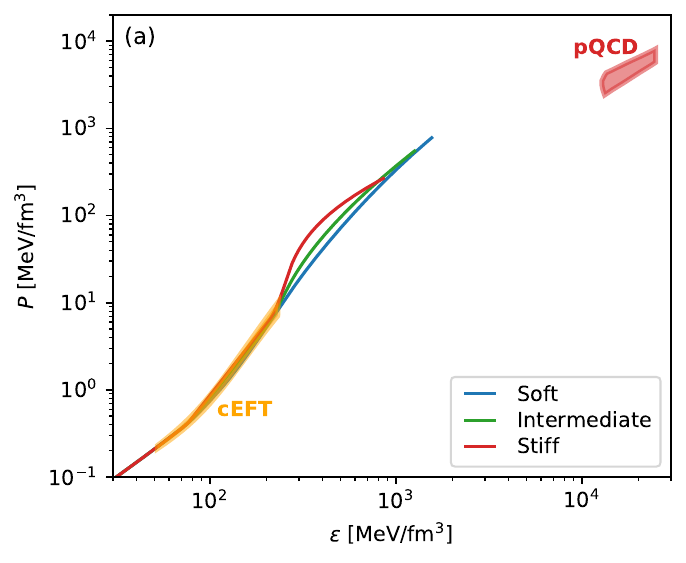}
\includegraphics[height=.353\linewidth]{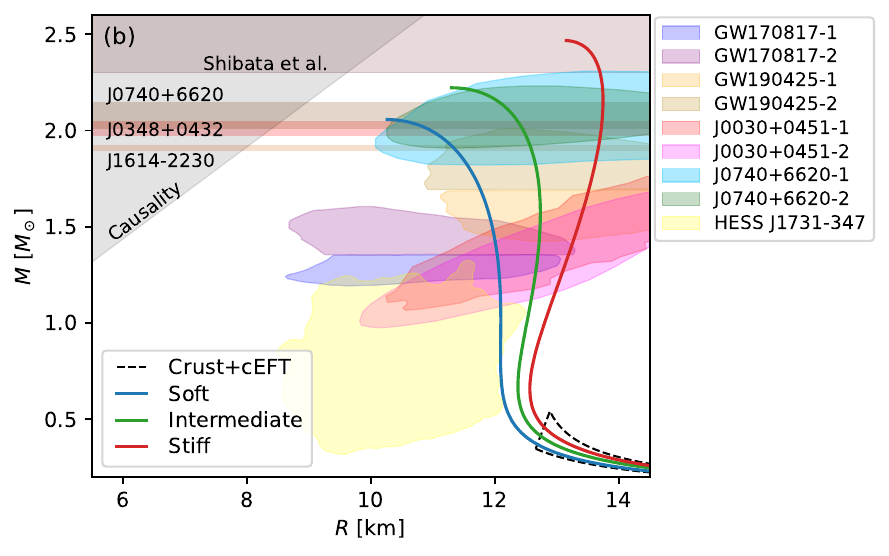}
\caption{$P$-$\epsilon$ and $M$-$R$ relationship for the \soft, \inter, and \stiff hadronic EoSs. (a) In the $P$-$\epsilon$ plane, the EoSs are presented up to the energy density value reached at the center of the maximum mass NS configuration. Constraints are provided by cEFT and pQCD as presented in Ref.~\cite{Drischler:2021lma} and Ref.~\cite{Annala:2020efq}, respectively. (b) In the $M$-$R$ plane we also show astrophysical constraints from the \mbox{$\sim 2~M_\odot$} pulsars \cite{Demorest:2010sdm, Antoniadis:2013amp, Arzoumanian:2018tny, Cromartie:2020rsd, Fonseca:2021rfa}, GW170817 \cite{Abbott:2017oog, Abbott:2018gmo} and GW190425 \cite{Abbott:2020goo} events ($90\%$ posterior credible level), NICER pulsars ($2\sigma$ posterior credible level) \cite{Riley:2021anv, Miller:2021tro,Miller:2019pjm, Riley:2019anv}, and HESS~J1731-347 ($2\sigma$ posterior credible level) \cite{Doroshenko:2022asl}. The upper horizontal area is the region excluded by \citet{Shibata:2019ctb}, $M_\textrm{max} \leq 2.3~M_\odot$. The shaded region in the upper left corner indicates the causality forbidden zone. The dashed black curve shows the TOV integrated region of the cEFT constraint. These hadronic EoSs represent extreme limiting cases for both nuclear and astrophysical constraints (see text for details).}
\label{fig:eos-mr_had}
\end{figure*}

In this section, we outline the criteria used for constructing the hadron and quark EoSs, as well as  the phase transition associated with the hybrid EoS. To make our study of HSs comprehensive and independent of specific hadron and quark EoSs models, we consider generic parametric forms of the hybrid EoSs. These forms allow us to fit not only the constraints imposed by cEFT but also those arising from pQCD, representing a diverse range of microscopic hybrid EoSs.

To construct the hadron sector of the hybrid EoS, we follow the prescription of Ref.~\cite{OBoyle:2020peo}, as  previously done in Refs.~\cite{lugones:2023ama, Ranea:2022bou, Ranea:2023cmr, Ranea:2023auq}, and consider the BPS-BBP crust \cite{Baym:1971tgs, Baym:1971nsm}. We use the BPS-BBP crust up to $0.5~n_0$, and beyond this density, we apply the GPP prescription, ensuring continuity in $n$, $P$, and $\epsilon$ at the juncture. The junction density $0.5~n_0$ corresponds to the value $\log_{10}(\rho_0) = 14.127$, marking the beginning of the first section of the piecewise polytropic EoS. 

In this way, we have constructed three representative hadronic EoSs, labeled  as \soft, \inter and \stiff. The detailed parameters of each hadron EoS are presented in Table~\ref{tabla:eos_had}. The choice of the limiting \soft and \stiff EoSs allow us to span a wide range of hadronic EoSs  consistent with cEFT. The inclusion of an \inter EoS aims to enrich our study and broaden its scope. In Fig.~\ref{fig:eos-mr_had}(a), we show their $P$-$\epsilon$ relationship along with current constraints from cEFT \cite{Drischler:2021lma} and pQCD \cite{Kurkela:2009gj, Gorda:2018gpy, Annala:2020efq}; in the right panel, we show the correspondent $M$-$R$ relationships along with current astrophysical constraints. It can be observed how \soft and \stiff EoSs act as enveloping curves for this low-density cEFT constraint (considered up to $1.5~n_0$, as model uncertainties tend to increase for larger densities); the \inter EoS is in between of them, in both the $P$-$\epsilon$ and $M$-$R$ planes. Moreover, by construction, the \soft hadronic EoS satisfies all current astronomical constraints by its own (in agreement with some previous works mentioned in Sec.~\ref{sec:intro}); the \inter one satisfies all current astronomical constraints except the one associated with HESS~J1731-347; the \stiff one satisfies all current astronomical constraints except those from HESS~J1731-347 and GW170817. Additionally, the \stiff EoS does not satisfy the \mbox{$M_{\max} \lesssim 2.3~M_\odot$} condition either \cite{Rezzolla:2018ugw, Shibata:2019ctb, Musolino:2024otm}. However, in the HS TOV integration, the appearance of the quark core  will allow it to satisfy the maximum mass upper constraint. In this sense, \soft and \stiff EoSs act as limiting extreme cases also in the $M$-$R$ plane, representing the frontier zone of the minimum and maximum allowed radius and masses, respectively.

\begin{table}
\begin{tabular}{cccccccc}
\toprule
 & $\log_{10}\rho_1$ & $\log_{10}\rho_2$  & $\Gamma_1$ & $\Gamma_2$ & $\Gamma_3$ \vspace{0.1cm}  \\
\midrule
\soft &  14.30 & 14.65 & 2.757 & 3.4 & 2.70 \\
\inter & 14.50 & 14.65 & 2.763 & 5.1 & 2.35 \\
\stiff & 14.55 & 14.65 & 2.766 & 9.5 & 1.05 \\
\bottomrule
\end{tabular}
\caption{Parameters of the selected hadronic EoSs constructed using the prescription of Ref.~\cite{OBoyle:2020peo} and the BPS-BBP crust \cite{Baym:1971tgs, Baym:1971nsm} up to $0.5~n_0$. In all cases, we adopted $\log_{10}(\rho_0) = 14.127$ and $\log_{10}(K_1) = -27.22$ because the crust ends at $0.5~n_0$. The EoSs are also consistent with the boundaries of the region allowed by cEFT presented in Ref.~\cite{Drischler:2021lma}, as shown in Fig.~\ref{fig:eos-mr_had}(a).}
\label{tabla:eos_had}
\end{table}

For the quark sector, we consider the constant speed of sound (CSS) model \cite{alford:2015caa}. This model is a general parameterization used to represent qualitative aspects of various models of quark matter, suitable for high-density EoS and characterized by a first-order abrupt hadron-quark phase transition, where the speed of sound in the quark phase is independent of pressure. The model consists of three parameters: the hadron-quark transition pressure, $P_t$, the size of the energy density transition jump, $\Delta \epsilon$, and the squared speed of sound of the quark phase, $c_s^2$, which is constant within the model. 

Both the parametric models for hadron and quark matter enable us to extensively explore the hybrid EoS space, aiming to obtain general qualitative and model-independent results. Thus, considering an abrupt first-order hadron-quark transition, we have constructed
\begin{equation*}
    3 \ \mathrm{(hadron)} \times 13 \ (P_t) \times 13 \ (\Delta \epsilon) \times 3 \ (c_s^2) = 1521
\end{equation*}
hybrid EoSs using the three \soft, \inter and \stiff hadronic EoSs, and the following parameter ranges for the CSS quark EoS:
\begin{eqnarray*}
    10~\mathrm{MeV/fm}^3 \le &P_{t}& \le 300~\mathrm{MeV/fm}^3 \, , \\
    100~\mathrm{MeV/fm}^3 \le &\Delta \epsilon& \le 3000~\mathrm{MeV/fm}^3 \, , \\
    c_s^2 &=& 0.33, 0.50, 0.70 \, .
\end{eqnarray*}
All of the $1521$ constructed EoSs satisfy the cEFT constraint. Only the EoS with $c_s^2 = 0.33$ fulfill the pQCD constraint, while the EoSs with $c_s^2 = 0.50, 0.70$ do not (see light gray curves in Fig.~\ref{fig:eos-mr}(a)). However, we do not discard them yet, as they remain valid if the central density and pressure of the last stable NS configuration do not exceed the pQCD densities and pressures, as we will discuss later.

It is crucial to mention at this point that the speed of the conversion between hadronic and quark matter at the phase transition surface plays a key role in the stability analysis of HSs \cite{Pereira:2017rmp} (see also \cite{lugones:2021pci} for a review on the subject). First-order phase transitions are highly collective non-linear phenomena and, for this reason, the conversion timescale between phases might not conform to the typical interaction timescales. Nucleation is one of the preferred mechanisms to model such processes (see, for example, Ref.~\cite{bombaci:2016qmn} and references therein). It is important to recall that alternative theoretical approaches are available in the literature (see, for example, Refs.~\cite{olinto:1987oct,alford:2015pcd} and references therein). Within the nucleation scenario, a direct conversion between hadronic and quark phases in $\beta$ equilibrium is suppressed as the phases have, in principle, very different flavor compositions. For this reason a two-step conversion mechanism has been proposed \cite{Lugones:1997gg, Lugones:2015bya, bombaci:2016qmn}. This establishes the slow conversion regime (where conversion timescales may exceed a few milliseconds) as a plausible scenario for a mechanism not yet fully understood (for a more detailed discussion, see the recent review \cite{lugones:2021pci} and references therein). In this scenario, a (long) branch of SSHSs with $\partial M / \partial \epsilon _c < 0$ can be present after solving the TOV equations that determine stationary stellar configurations \cite{mariani:2019mhs,malfatti:2020dba, curin:2021hsw, rodriguez:2021hsw, Mariani:2022omh,Goncalves:2022ios, Rau:2023neo,Rau:2023tfo,Rather:2024roo}. For this reason, to construct the abrupt phase-transition, we work within the \textit{slow} hadron-quark conversion scenario, which gives rise to SSHSs configurations located on an extended stability branch following the maximum mass configuration in the $M$-$R$ relationship. This contrasts with the traditional TSHSs that are stable in both the \textit{rapid} and \textit{slow} scenarios along the $\partial M / \partial \epsilon _c > 0$ branches.

\section{Results}
\label{sec:results}

\begin{figure*}[t]
\centering
\includegraphics[height=0.411\linewidth]{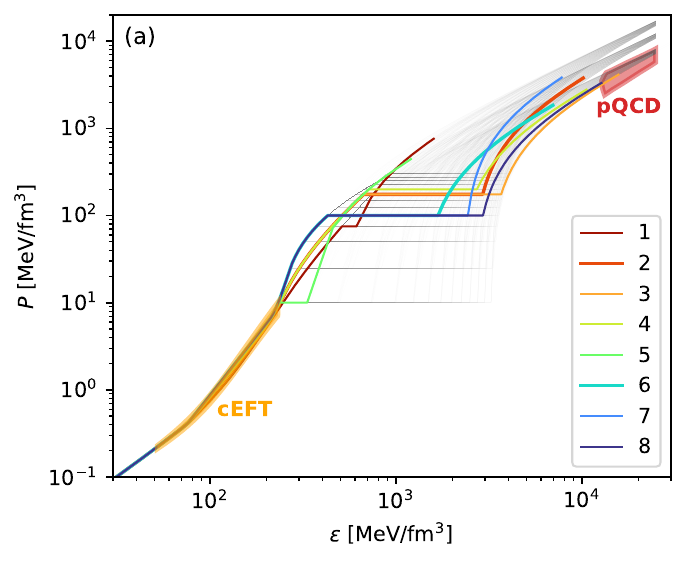}
\includegraphics[height=0.411\linewidth]{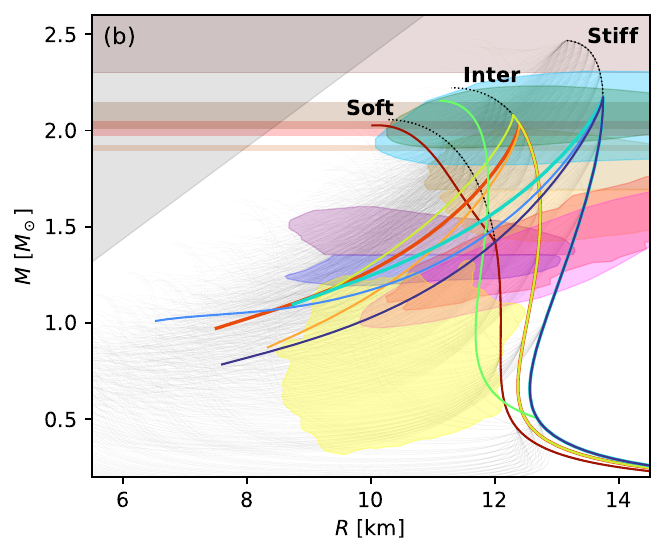}
\caption{(a) The eight selected hybrid EoSs are shown in color, along with the cEFT constraint up to $1.5\, n_0$ \cite{Drischler:2021lma} and pQCD results for $n \gtrsim 40\, n_0$ \cite{Kurkela:2009gj, Gorda:2018gpy, Annala:2020efq}. The selected EoSs are displayed up to the largest central density reached in each HS model. Details of the EoSs are provided in Table~\ref{tabla:eos_selec}. All 1521 constructed hybrid EoSs, up to the largest calculated density, are shown in light gray. (b) Mass-radius relationships for the EoSs in the panel (a); the same color-coding is used for the eight selected cases.  The fully hadronic \soft, \inter, and \stiff curves are shown in black as a reference. Since SSHSs are considered, all curves are shown up to the terminal configuration, and thus all points of the curves represent dynamically stable stars. The colored shaded regions represent the constraints described in Fig.~\ref{fig:eos-mr_had}. The TOV results for all 1521 hybrid EoS are presented in light gray.}
\label{fig:eos-mr}
\end{figure*}

Given the extensive sampling of hybrid EoSs constructed in the previous section, we select eight of them that satisfy all current nuclear and astrophysical constraints, including HESS~J1731-347. In Table~\ref{tabla:eos_selec} we present the details of these EoSs, which exhibit different hadronic components  and a wide range of the CSS parameters. We selected them to demonstrate the variety in qualitative EoS behaviors and the lack of need of fine tuning. It is important to note that the selected EoSs serve as representatives of EoS families with common qualitative morphology that satisfy the constraints, as numerous sampled EoSs meet these criteria beyond the particular eight selected.  Additionally, we aimed to include EoSs that are potentially limiting or extreme cases, to serve as enveloping samples for intermediate cases (see the discussion below regarding EoSs~$2$ and $7$).

\begin{table}
\begin{tabular}{ccccc}
\toprule
 \# & Hadronic & $P_t$[MeV/fm$^3$]  & $\Delta \epsilon$[MeV/fm$^3$] & $c_s^2$ \vspace{0.1cm}  \\
\midrule
1 & \soft &  75 & 100 & 0.70 \\
2 &\inter & 175 & 2250 & 0.50 \\
3 &\inter & 175 & 3000 & 0.33 \\
4 &\inter & 200 & 2000 & 0.33 \\
5 & \stiff & 10 & 100 & 0.50 \\
6 & \stiff & 100 & 1250 & 0.33 \\
7 & \stiff & 100 & 2000 & 0.70 \\
8 & \stiff & 100 & 2500 & 0.33 \\
\bottomrule
\end{tabular}
\caption{Parameters of the eight selected EoSs among the 1521 constructed hybrid EoSs (see text for more details).}
\label{tabla:eos_selec}
\end{table}

In  Fig.~\ref{fig:eos-mr}(a), we present the results for the selected hybrid EoSs, along with all 1521 hybrid EoSs constructed, show as light gray curves in the background. In addition to the the $P$-$\epsilon$ relationships, we include the cEFT and pQCD constraints. The EoS curves depict the hadron phase for lower densities up to the abrupt phase transition indicated by the energy density jump, and the quark sector for higher densities. These EoSs are presented up to the maximum energy density reached in the stable NS core.  As shown, for the cases with $c_s^2=0.50, 0.70$, the physically relevant energy density and pressure ranges do not exceed --and are thus compatible with-- the pQCD prediction. In particular, EoSs $2$ and $7$ present a limiting case in terms of pQCD compatibility: we select these two to sample these extreme yet possible cases where a sequential second phase transition could occur in the extreme high-density regime. The other selected EoSs automatically satisfy pQCD as they have $c_s^2=0.33$ (EoSs $3$, $4$, $6$ and $8$) or they do not reach such high densities in the NS cores and could potentially exhibit a smooth decrease in the speed of sound (EoSs $1$ and $5$). The numerous gray curves show that, for large densities, the EoSs converge into three distinctive families corresponding to $c_s^2 = 0.33, 0.50, 0.70$, and that only $c_s^2 = 0.33$ satisfies pQCD for these densities.

In Fig.~\ref{fig:eos-mr}(b) we show the $M$-$R$ relationships for the selected EoSs. Similar to Fig.~\ref{fig:eos-mr}(a), we include all $1521$ TOV results as light gray curves in the background. Each configuration shown in the figure is dynamically stable within the slow conversion regime (i.e., after solving the radial perturbation equations, we find that their fundamental radial eigenfrequency is real until the \textit{terminal mass}, at which point it vanishes). We can see that every astronomical constraint is satisfied by these representative cases. EoS $1$ and $5$, with the the lowest transition pressures, become hybrid for relatively low masses, exhibit short extended branches, and satisfy the constraints with their TSHS branches. EoS~$1$ corresponds to a \soft hadronic EoS (satisfying the constraints without any hadron-quark transition); a high speed of sound in the quark sector, $c_s^2=0.70$, allows the hybrid branch to extend beyond $2.01~M_\odot$, satisfying all constraints even with the occurrence of the hadron-quark phase transition. For this EoS, the HESS~J1731-347 object would be a hadronic NS, while GW170817 objects could be both hadronic and/or hybrid configurations. On the other hand, EoS~$5$ corresponds to a \stiff hadronic EoS, and the very early phase transition, along with decreasing radii, enables it to satisfy HESS~J1731-347 and GW170817 constraints through a TSHS branch with very low masses. As already shown in \citet{lugones:2023ama}, these two EoSs confirm that the low-pressure transition scenario remains valid for meeting modern astrophysical constraints. However, the kind of compact objects produced by these two EoSs are well-studied \cite{Kubis:2023rmf,Sagun:2023wit,Laskos:2024hsi}, do not involve the SSHS branch, and are not the primary focus of our work. The remaining six EoSs are selected to show that the extended SSHS branch is a possible alternative model for simultaneously meeting the HESS~J1731-347 and \mbox{PSR 0740+6620} constraints. These six EoSs produce a stable hadronic branch with large radii up to a maximum mass configuration. At this point, the hadron-quark phase transition occurs, followed by a steep and extended SSHS branch that overlaps with the HESS~J1731-347 measurement, roughly at $1~M_\odot$ and within the $9$-$11$~km range.

Although the \soft EoS is not the optimal candidate for producing SSHSs that satisfy current constraints due to its low maximum mass and small radii, both the \inter and the \stiff EoSs offer an extended parameter space that meets all requirements. Our results suggest that the CCO in HESS~J1731-347 could be either a purely hadronic NSs or a SSHSs. In particular, they indicate that the HESS~J1731-347 object could be a particularly compact and low mass SSHS, provided the hadronic EOSs lies between the \inter and \stiff cases. As our generic proposals are constructed to be model-independent limiting cases,  numerous microphysical EoS models in the literature likely fall within these bounds and could inherit the implications of our findings.

\begin{figure*}[t]
\centering
\includegraphics[height=.352\linewidth]{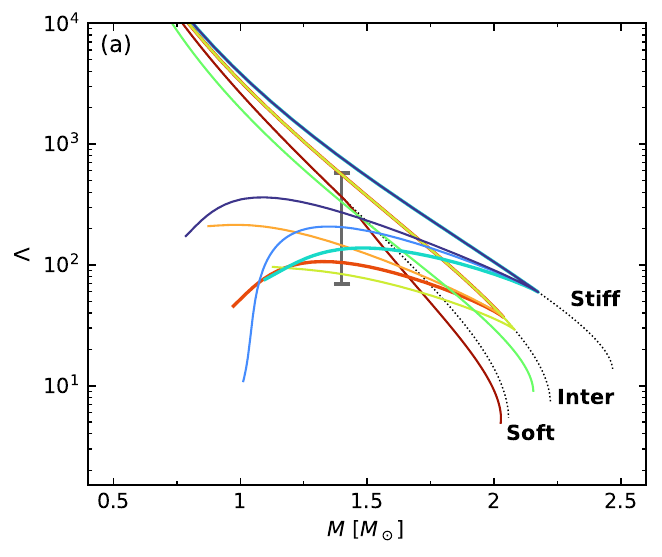}
\includegraphics[height=.352\linewidth]{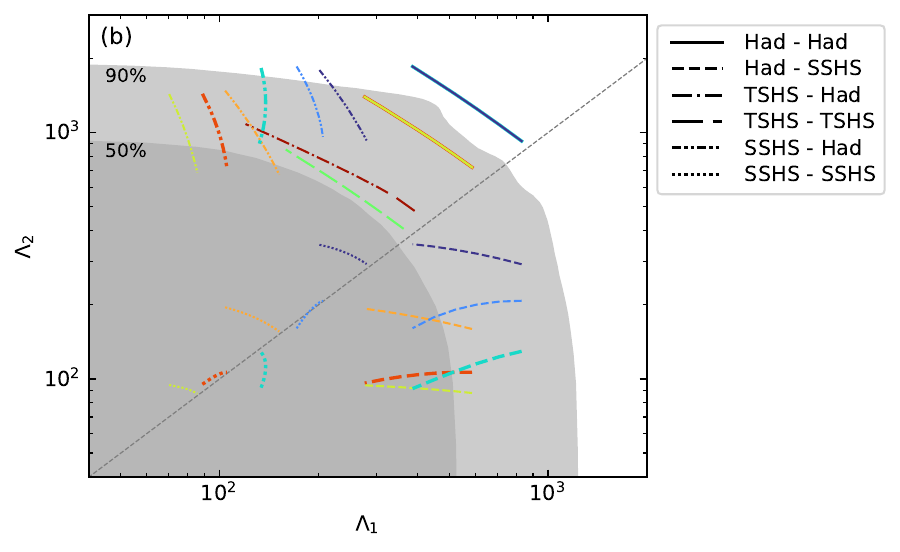}
\caption{(a) Dimensionless tidal deformability, $\Lambda$, as a function of the gravitational mass, $M$, for the EoSs of Fig.~\ref{fig:eos-mr} (same colors are used). Since we are considering slow hadron-quark conversions and SSHSs extended branches, all the shown curves represent dynamically stable configurations. The fully hadronic \soft, \inter, and \stiff curves are shown in black as a reference. The gray vertical segment represents the constraint imposed for a $1.4 \, M_\odot$ NS by the analysis of the GW170817 event \cite{Abbott:2018gmo}. (b) Individual tidal deformabilities $\Lambda_1$, $\Lambda_2$ for binary neutron star mergers with the same chirp mass as the GW170817 binary system \cite{Abbott:2018exr, Abbott:2018pot, Abbott:2018gmo} for the EoSs of Fig.~\ref{fig:eos-mr} (same colors are used). Each line type indicates the kind of objects potentially involved in the binary merger. For slow conversions, $\Lambda_1 > \Lambda _2$ is possible because $\Lambda$ does not necessarily decrease with $M$. Dark and light gray areas represent the $50\%$ and $90\%$ confidence levels of the GW170817 event, respectively.}
\label{fig:tidal}
\end{figure*}

In Fig.~\ref{fig:tidal}(a) we present our results for the dimensionless tidal deformability, $\Lambda$, in the $\Lambda$-$M$ plane, along with the constraint arising from the GW170817 event for a $1.4 \, M_\odot$ configuration \cite{Abbott:2018gmo}. Similar to the $M$-$R$ results, EoSs $1$ and $5$ satisfy this constraint through the totally stable branch, purely hadronic for EoS $1$ and hybrid for EoS $5$. The other EoSs with long SSHS branches satisfy this constraint through the extended stability branch, which exhibits a non-monotonic behavior of $\Lambda$ as a function of $M$, with the $1.4 \, M_\odot$ configuration being a SSHS. In Fig.~\ref{fig:tidal}(b), we show the individual dimensionless tidal deformabilities in the $\Lambda_1$-$\Lambda_2$ plane, considering the stellar configuration pairs that match the chirp mass of the GW170817 binary neutron star merger and the $50\%$ and $90\%$ confidence contours for this event (both assuming low-spin priors) \cite{Abbott:2018exr,Abbott:2018pot,Abbott:2018gmo}. In this figure, there are numerous disconnected branches for each EoS as different types of compact object combinations are considered. The region where $\Lambda_1 > \Lambda_2$ is populated since in the \textit{slow} conversion stability scenario, $\Lambda$ can increase with $M$. Every EoS has at least one branch combination that falls inside the confidence regions. Combinations including SSHS-type objects are statistically preferred as they mostly fall inside the $50\%$ region, compared to purely hadronic and/or TSHS pairs.

\section{Summary and discussion}
\label{sec:conclus}

In this work, we provided an alternative explanation for the CCO in the supernova remnant HESS~J1731-347. For the first time, we showed that SSHSs are viable candidates to explain the observed mass and radius of this ultra-compact star using a model-agnostic approach to construct hybrid EoSs.

We constructed $1521$ hybrid EoSs satisfying the cEFT constraint using parametric EoSs suitable for the crust-hadron and quark phases: the generic BPS-BBP-GPP and CSS models, respectively. Eight of them were selected to represent a variety of EoSs with qualitative different behaviors without requiring fine-tuning, while still satisfying all current nuclear and astrophysical constraints, including those from HESS~J1731-347. In selecting the EoSs, we also took into account results from pQCD calculations. This constraint is fulfilled by EoSs $3$, $4$, $6$, and $8$ by construction ($c_s^2=0.33$); by EoSs $2$ and $7$ if there is a second abrupt phase transition at very high density; and by EoSs $1$ and $5$ if a smooth decrease in the speed of sound is assumed beyond the largest NS density. Our results show that the characteristics of the HESS~J1731-347 object can be explained by some models of pure hadronic stars and traditional hybrid stars. However, this is only possible for very specific hadronic models (soft hadronic EoS) or traditional hybrid models with a phase transition at very low density (EoSs 1 and 5). In contrast, our results demonstrate that there is a wide variety of SSHS models that naturally satisfy the HESS~J1731-347 observation.

Regarding the dimensionless tidal deformability $\Lambda$ and the $\Lambda_1$-$\Lambda_2$ relationship, the obtained results are similar to those of Ref.~\cite{lugones:2023ama}: in the context of SSHSs, the most massive component of a binary compact object merger would not necessarily have the smallest $\Lambda$. Additionally, in line with the findings of the aforementioned work, binary systems involving at least one object from the SSHS branch rather than only combinations of the traditional hadronic and/or TSHS ones, could be considered as an alternative explanation for the GW170817 event.

Our model for HESS~J1731-347 avoids some potential issues that other theoretical proposals might encounter. For instance, the ultra-soft hadronic EoS needed to provide a (non-marginal) explanation for HESS~J1731-347 as a purely hadronic NS are in tension with modern results from cEFT calculations. As discussed and summarized in Refs.~\cite{Lope:2019nns, Lope:2024cpg, Sorensen:2024dnm}, various modern cEFT calculations for the low-density region of the hadronic EoS exist (e.g., Refs.~\cite{Hebeler:2013nza, Lynn:2016ctn, Hu:2017nmp, Holt:2017eos, Drischler:2020hwd}); in this study, as previously mentioned, we applied the constraints presented in \citet{Drischler:2021lma}. However, any of the more recent cEFT constraints strongly restrict the low-density EoS regime, requiring even more careful tuning of the hadronic model to match the results of HESS~J1731-347.

On the other hand, concerns have been raised about the idea that HESS~J1731-347 might be a SQS, mainly because observations indicate that the star's surface temperature is quite high \cite{Doroshenko:2022asl}. This would conflict with the rapid cooling expected in quark matter due to the onset of the quark direct URCA process (see, for example, Refs.~\cite{blaschke:2001coh, wei:2020coh, zapata:2022tra} and references therein). However, a viable explanation for the surface temperature of \mbox{HESS~J1731-347} could be achieved if quarks form color superconducting pairs with a small gap, which suppresses the fast cooling \cite{Horvath:2023als, Sagun:2023wit}. The primary obstacle of this interpretation lies in the inherently speculative nature of the strange quark matter hypothesis, which requires not only the formation of deconfined quark matter in compact objects but also that such matter be self-bound.

On the other hand, models suggesting that the HESS~J1731-347 object comprises normal matter admixed with dark matter require a dark matter fraction of approximately $5\%$ of the total NS mass to account for the observed mass and radius of this object (see, e.g., Ref.~\cite{Sagun:2023wit} and references therein). Considering that the local dark matter mass density in the solar neighborhood is estimated to be around $0.01 \, M_{\odot} \, \mathrm{pc}^{-3}$ \cite{Catena:2009and}, the capture efficiency of dark matter particles appears exceedingly low \cite{Kouvaris:2007ay}, making it difficult to justify how such a significant amount of dark matter could accumulate inside the compact object in HESS~J1731-347, which is only a few $\mathrm{kyr}$ old.

To conclude, both the precise characteristics of the HESS~J1731-347 compact object and the existence of SSHSs (given the ongoing debate surrounding the nature of the hadron-quark phase transition) remain open questions. If SSHS are indeed possible, and if future observations were to confirm the lowest mass NS to date within HESS~J1731-347, SSSHs could provide a plausible explanation for this compact object and shed light on the still unknown EoS of NSs.

\section*{Acknowledgments}
M. M. thanks C. Drischler for providing the data constraints for cEFT. M.M, I.F.R-S, and M.G.O acknowledge UNLP and CONICET for financial support under grants G187 and PIP-0169. M.M is a postdoctoral fellow of CONICET. I.F.R-S is also partially supported by PIBAA 0724 from CONICET. I.F.R-S and M.G.O. are partially supported by the National Science Foundation (USA) under Grant PHY-2012152. G.L. acknowledges the financial support from  CNPq (Brazil) grant 316844/2021-7 and FAPESP (Brazil) grant 2022/02341-9. 

\bibliography{references}

\end{document}